\documentclass{aa}  
\usepackage{todonotes}
\usepackage{graphicx}
\usepackage{amsmath}
\usepackage[flushleft]{threeparttable}
\usepackage{txfonts}

\newcommand{\sga}{Sgr~A$^*$}

\newcounter{todocounter}
\newcommand{\greg}[1]{\stepcounter{todocounter}
  {\color{red!90} Greg: \thetodocounter: #1}}

\newcommand{\ralph}[1]{\stepcounter{todocounter}
  {\color{gray!90} Ralph: \thetodocounter: #1}}

\begin{document}

   \title{A survey for radio pulsars and transients in the 10\,pc region around Sgr~A$^{*}$}


   \author{G.~Desvignes\inst{1}\fnmsep\thanks{Email: gdesvignes@mpifr-bonn.mpg.de}
          \and
          R.~P.~Eatough\inst{2,1}
          \and
          Y.~Men\inst{1}
          \and
          F.~Abbate\inst{3,1}
          \and
          R.~Karuppusamy\inst{1}
          \and
          M.~Kramer\inst{1,4}
          \and
          K.~Liu\inst{5,6,1}
          \and
          L.~Shao\inst{7,1,8}
          \and
          P.~Torne\inst{9}
          \and
          R.~S.~Wharton\inst{1}
          }

   \institute{Max-Planck-Institut f\"{u}r Radioastronomie, Auf dem H\"{u}gel 69, D-53121 Bonn, Germany
     \and
     National Astronomical Observatories, Chinese Academy of Sciences, 20A Datun Road, Chaoyang District, Beijing 100101, People’s Republic of China
     \and
     INAF – Osservatorio Astronomico di Cagliari, Via della Scienza 5, 09047 Selargius (CA), Italy
     \and
     Jodrell Bank Centre for Astrophysics, Department of Physics and Astronomy, The University of Manchester, Manchester M13 9PL, UK 
     \and
     Shanghai Astronomical Observatory, Chinese Academy of Sciences, 80 Nandan Road, Shanghai 200030, China
     \and
    State Key Laboratory of Radio Astronomy and Technology, A20 Datun Road, Chaoyang District, Beijing, 100101, P. R. China
    \and
     Kavli Institute for Astronomy and Astrophysics, Peking University, Beijing 100871, People’s Republic of China
     \and
     National Astronomical Observatories, Chinese Academy of Sciences, Beijing 100012, People’s Republic of China
     \and
     Instituto de Radioastronomía Milimetrica, Avda. Divina Pastora 7, Núcleo Central, E-18012 Granada, Spain
             }

   \date{}

 
  \abstract
  {Here we report on a new survey for pulsars and transients in the 10\,pc region around \sga{} using the Effelsberg radio telescope at frequencies between 4 to 8 GHz. Our calibrated full-Stokes  data were searched for pulsars and transients using \textsc{PulsarX}, \textsc{TransientX} and \textsc{PRESTO}. Polarisation information is used in the scoring of the candidates. Our periodicity acceleration and jerk searches allowed us to maintain good sensitivity towards binary pulsars in $\gtrsim 10$-hr orbits. In addition we performed a dedicated search in linear polarisation for slow transients. 
  While our searches yielded no new discovery beyond the redetection of the magnetar SGR~J1745$-$2900, we report on a faint single pulse candidate in addition to several weak periodicity search candidates. After thoroughly assessing our survey's sensitivity, we determined that it is still not sensitive to a population of millisecond pulsars. Next generation radio interferometers can overcome the limitations of traditional single-dish pulsar searches  of the Galactic Centre.}
  


   \maketitle
%

\section{Introduction}

The timing of a radio pulsar in a sufficiently close orbit (orbital
period $P_{\rm b}\lesssim 1$\,yr) around Sagittarius~A* (\sga{}) - the supermassive black hole at the heart of our Galaxy - promises unrivaled tests of
gravitational physics \citep{lwk+12} and would allow for complimentary
constraints on the results from the Event Horizon Telescope (EHT) and the GRAVITY experiment at the Very Large Telescope Interferometer \citep{pwk16}. 
It was demonstrated recently that pulsars with slightly larger orbital periods, $P_{\rm b} \sim 2 \mbox{--} 5 \, {\rm yr} $, are also useful to measured the black hole spin to good precision \citep{hs24}.

For these reasons, the Galactic Centre (GC) has been the subject of numerous pulsar surveys over the past few decades. 
Despite this \citep[see e.g.][]{kkl+00,kle05,mkf+10}, only five slow pulsars 
were found in the GC region until 2013; 
all with a projected distance $>25$\,pc from \sga{} \citep{jkl+06,dcl09}.


\citet{wcc+12} predicted up to about 100 canonical pulsars (i.e. with spin period $P_\textrm{s}$ of order 0.1 - 1 s ) and 1000 recycled pulsars (i.e. with spin $P_\textrm{s} \lesssim 30$\,ms ) in the central parsec of the GC.
The GC 
could also harbour several recycled pulsars  in binary orbits with stellar mass black hole companions \citep{fl11}.
As described above, none of the past surveys detected a pulsar 
closer than 25\,pc in projection
from \sga{}.  Nominally ``hyper-strong'' scattering towards the GC
was 
blamed for this lack of detection, with an estimated
scattering time $\tau_{1\textsc{GHz}}$ of about 1000\,s at 1\,GHz \citep{cl02}. Assuming a
Kolmogorov spectrum with spectral index $\alpha=-3.8$, the scattering
time would scale as $\tau \sim f^\alpha$ prohibiting any detection of
a millisecond-duration pulsed signal at frequencies below $\sim $10\,GHz.

This explanation 
has come under debate
since the 
discovery of an
X-ray outburst and pulsations from the magnetar
SGR~J1745--2900  \citep{kbk+13}  
and eventually the detection of radio pulsations just a few days later
\citep{efk+13}. With a projected distance of only 2.4 arcsec \citep{rep+13}, a
dispersion measure (DM; the integrated free electron column density
along the line of sight) of 1778\,pc\,cm$^{-3}$ and a rotation measure
(RM; the rotation in the plane of polarisation of emission due to an external magnetic field) of $-66960$\,rad\,m$^{-2}$ at the time of discovery \citep{efk+13}, this magnetar is thought to be within the Bondi-Hoyle accretion limit of \sga{}.
Measurements of the 
scattering time of
SGR~J1745--2900's pulsed
emission \citep{sle+14} 
revealed an unexpectedly small
scattering
time of
$\tau_{1\textsc{GHz}}=1.3\pm0.2$\,s, compared to 
the hyper-strong scattering 
predicted by the NE2001 model \citep{cl02}. This result
potentially allows for the detection of a non-fully-recycled
millisecond pulsar (MSP) in the GC at frequencies $>4$\,GHz, with
corresponding scatter broadening time of $\lesssim 7$ ms.

The question arises then: 
why haven't more pulsars with $P_\textrm{s}\lesssim1$\,s been found within the 25\,pc region around \sga{} despite more recent surveys with increased sensitivity to both canonical and recycled pulsars \citep{wha17,scc+21,tde+21,etd+21,tle+23}?
\citet{do14} interpreted this discrepancy as possible evidence that the massive stars in the GC predominantly form magnetars instead of canonical pulsars.

We report here on new observations of the GC region (\S2) intended to better constrain the GC pulsar population. We describe our search pipeline for pulsar and transients in \S3 and discuss the results of this work in \S4. We present our conclusions in \S5.

\section{Observations}
We used the 100-m Effelsberg radio telescope of the Max Planck Institute for Radio Astronomy  and its C-X band receiver,
with a 4 to 9.3\,GHz frequency coverage, to survey the projected 10\,pc region
around \sga{}. The PSRIX2 backend allowed us to record two contiguous
frequency bands of 2\,GHz with full Stokes information centered at 5 and 7\,GHz,
each split into 2048 frequency channels, with a sampling time $\delta t$ of
$131.072\,\mu$s. The two bands were combined offline with the data
converted into \textsc{PSRFITS}\footnote{https://www.atnf.csiro.au/research/pulsar/psrfits\_definition/Psrfits.html} search mode format and recorded with
8-bit values.

\begin{figure*}
    \begin{center}
      \includegraphics[width=\textwidth]{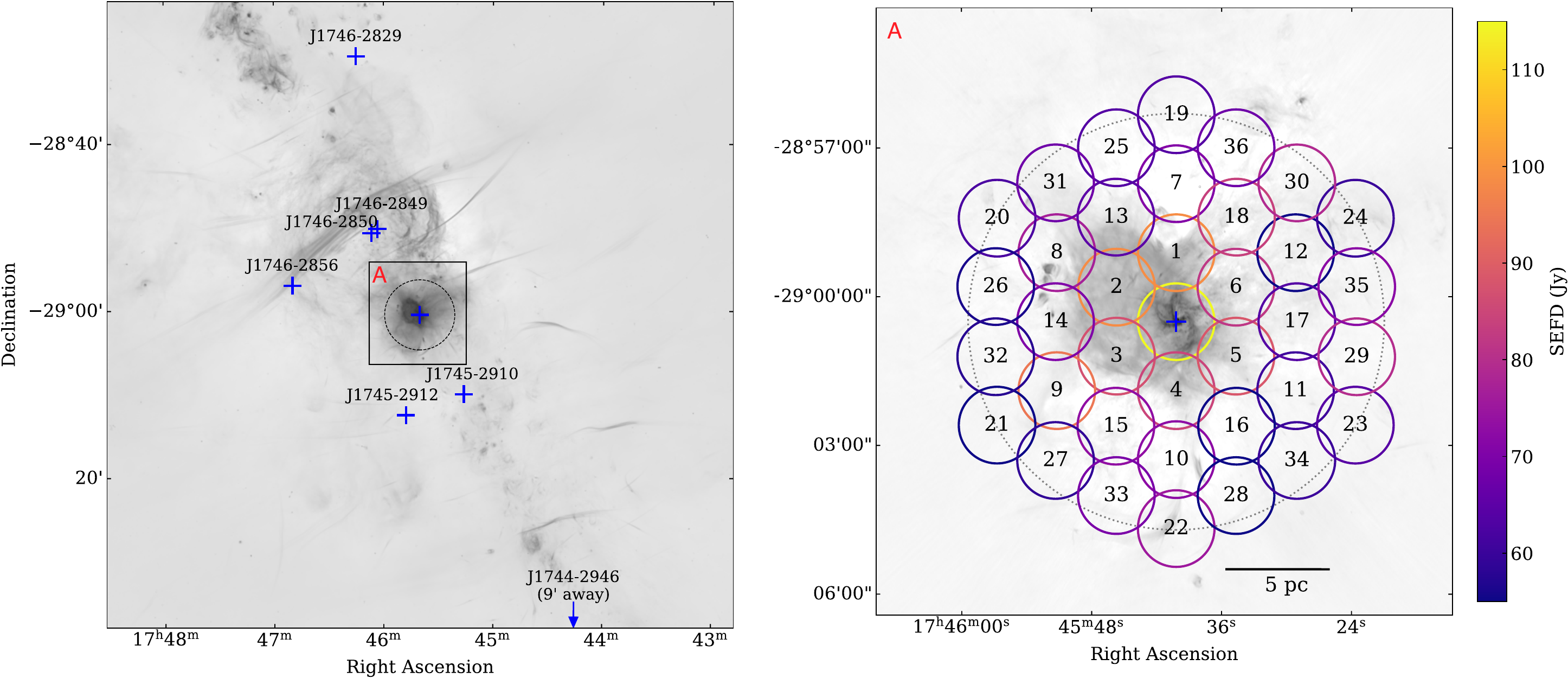}
        \caption{Left panel: Image of the GC region produced by MeerKAT at 1.28\,GHz \citep{hrc+22} with the positions of the known pulsars indicated by blue crosses (or a blue arrow in the case of the recently discovered millisecond pulsar J1744$-$2946, $9'$ away from the edge of the map). The region delineated by the square box marked with the letter A is displayed in the 
        right hand panel. The dashed grey circle delimits the 10\,pc region around \sga{} at the distance of the GC.  
        Right panel: Observing grid superimposed on a radio map of the GC obtained at 5.5\,GHz with the Very Large Array \citep{zmg16}. Each colored circle with the pointing number at its center represents the beam size at HPBW at 8\,GHz $\theta_{8\,\textrm{GHz}}=1.55'$  with the color corresponding to the pointing System Equivalent Flux Density (SEFD) value shown with the vertical colorbar. The blue cross at the center of the dotted circle indicate the location of the magnetar SGR J1745$-$2900. For clarity, we omit the numerical label of the inner pointing centered on \sga{}. The dotted grey circle delimits the 10\,pc region around Sgr~A*.}
    \label{fig:grid}
    \end{center}
\end{figure*}

We adopted the following strategy in our survey of the GC. Due to the
large fractional bandwidth of our observations, we took the half power
beam width (HPBW) of the Effelsberg telescope at 8\,GHz
(i.e. HPBW$_\textrm{8\,GHz} = 1.55'$) to tessellate our observing grid
around the inner pointing centred at the position of the GC magnetar
SGR~J1745$-$2900 ($\alpha_\textrm{J2000}=
17^{\textrm{h}}45^{\textrm{m}}40.190^{\textrm{s}}$,
$\delta_\textrm{J2000} = -29^\circ00'30''$) \citep{rep+13}. With the GC being at a distance $d_{\rm GC}=8.25\,\rm kpc$ \citep{grav+20}, three consecutive
tightly-bound rings of 6, 12 and 18 pointings with
HPBW$_\textrm{8\,GHz} = 1.55'$ 
cover the inner 10\,pc
around \sga{}, corresponding to a sky area of about 0.025 deg$^2$. The
observing grid is shown in Fig.\,\ref{fig:grid}.  Observations were
carried out between April 2019 and April 2020 and have an integration time $T_\textrm{int}$ of 1\,hr. Prior to each observation, we recorded a 2-min
scan with the pulsed noise diode on the same sky position to allow
for the polarimetric calibration of the data. Observations of the
noise diode fired at the sky location of the planetary nebula NGC~7027 (and also $1^\circ$ away from it) were scheduled monthly to assess the exact sensitivity of the survey (as discussed in Section \ref{sec:res}).
 

\begin{figure}
    \begin{center}
      \includegraphics[width=\columnwidth]{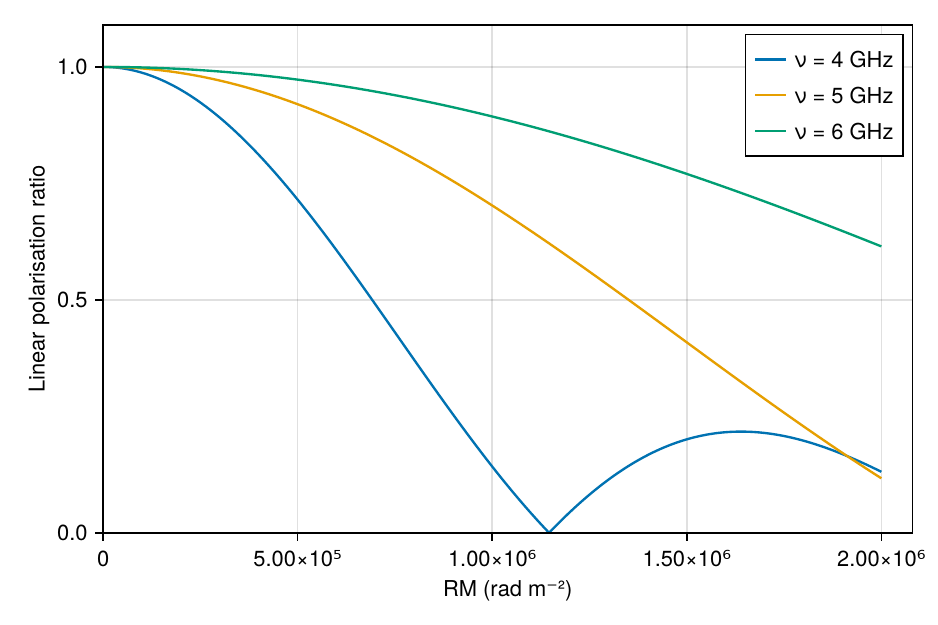}
        \caption{Polarisation ratio ($L_\textrm{obs}/L_\textrm{int}$) of the linear polarisation  as a function of RM for a frequency channel of width $\sim$0.976\,MHz at three reference frequencies $\nu$ of 4, 5 and 6\,GHz shown in blue, orange and green, respectively. $L_\textrm{obs}$ is the observed linear polarisation and $L_\textrm{int}$ is the instrinsic linear polarisation of the signal. The polarisation ratio is computed following \citet{sl15} and starts to show significant depolarisation across our channel bandwidth at 4\,GHz with RM $\gtrsim 500\,000$ rad\,m$^{-2}$. The RM polarisation ratio curve is symmetric around RM zero.}
    \label{fig:depol}
    \end{center}
\end{figure}

\section{Data processing}
We started by manually cleaning the noise diode observations from the effects of radio frequency interference (RFI) for the 37 pointings of our survey using \textsc{PSRCHIVE} \citep{sdo12}. On average, approximately 15\% of the bandpass was flagged in each observation, leading to an effective 3.4\,GHz bandwidth. Then we used the cleaned noise diode observation to calibrate in polarisation the PSRFITS search mode data. To do this \textsc{PSRCHIVE} is used to compute the Mueller matrix response for each frequency channel taking into account the effect of the parallactic angle and the differential phase and gain between the two linear feeds of the receiver. This matrix is then inverted and applied to the Stokes vectors recorded in the \textsc{PSRFITS} data \citep[for a review, see also e.g.][]{lk04}. This step requires floating-point arithmetic, and to avoid requantization, the calibrated full Stokes data are recorded with 32-bit float values. The frequency channels of the noise diode observation that were zapped due to RFI are zero-weighted in the PSRFITS calibrated output file and therefore discarded in the 
subsequent
analysis.

\subsection{Search in total intensity}
We used \textsc{PulsarX} \citep{mbc+23} to dedisperse the total intensity data (Stokes I) in the DM range from 0 to 5000\,pc\,cm$^{-3}$ with DM steps of 2\,pc\,cm$^{-3}$ while simultaneously applying the zero-DM and Kadane filters \citep{ekl09,mbc+23} for further RFI cleaning.
The search for periodicities in the dedispersed time series is performed in the Fourier domain with \textsc{PRESTO} \citep{rem02} in two passes: the first "low acceleration" pass allowing a frequency drift caused by a constant acceleration of up to 50 Fourier bins with up to 16 harmonics summed. The second pass is tuned to be more sensitive to non-linearly accelerated signals with up to 200 and 600 Fourier bins for the frequency drift and its derivative \citep[the "jerk" search,][]{ar18}, respectively. Due to limits in computing resources, we restricted the summation to 8 harmonics in the acceleration plus jerk search pass. Candidates for each pass are sifted separately to remove harmonically linked candidates or candidates with non-contiguous DM detections \citep[see for more details][]{lbh+15}.

The folding of the remaining candidates is also done with \textsc{PulsarX} after we implemented the handling of full Stokes data and folding with "jerk" information. For each folded candidate with a measured pulse window, we could then apply within \textsc{PulsarX} the Faraday rotation measure (RM) synthesis method \citep{bd05} to the detected pulse in the RM-range from $-100\,000$ to $+100\,000$ rad\,m$^{-2}$ with steps of 5 units of RM. Beyond this range, the intra-channel depolarisation in the lowest part of our frequency band becomes detrimental (see Fig.\,\ref{fig:depol}). The RM synthesis plot along with the RM value that maximise the amount of linear polarisation are added to the traditional \textsc{PulsarX} inspection plot of each candidate.
We restricted the folding of the candidates with \textsc{PulsarX} to the first 210 candidates (because of memory limits in the computing nodes) for both the low acceleration and the acceleration plus jerk search pass. All candidates produced were visually inspected.

In addition to the periodicity search, we performed a search for single pulses on all pointings with \textsc{TransientX}
\citep{mb24} with the same RFI cleaning and dedispersion parameters as described above. Similarly to our periodicity search, we modified \textsc{TransientX}  such that it could handle our full Stokes search data and perform RM synthesis on the detected single pulse candidate. We recorded all events with a signal-to-noise ratio (S/N) greater than 6 for visual inspection.

\subsection{Search for slow transients in linear polarisation}
\label{sec:ssearch}
Single-dish observations of the GC typically show a large amount of power fluctuations seen as red noise in the total intensity of the recorded time series due to, e.g. variations in atmospheric conditions and observed GC continuum emission \citep{etd+21,tde+21}. This justified our choice of using the zero-DM technique in the total intensity search described in the previous section, but, in turn, it decreases our sensitivity towards very wide pulses as the dispersion delay \citep{lk04} is limited to $<1$\,s across our frequency band with our maximum DM trial.

In the light of the recent discoveries of slow and highly linearly polarised transients with periods greater than tens of seconds \citep[e.g. ][]{chr+22,hzb+22}, we investigated a different approach to search for these very wide pulses in our data.
First, the \textsc{PSRFITS} data were time-scrunched by a factor of 128 after the polarisation calibration step, resulting in a new time resolution of $\sim 16.7$\,ms.  This allows us to limit the width of the boxcar used during the matched filtering process \citep{cm03} and drastically reduce the total number and length of time-series to be searched, making it more tractable with our current computing ressources. Then we dedispersed the Stokes data with DM steps of 200\,pc\,cm$^{-3}$ up to a maximum value of 5000\,pc\,cm$^{-3}$ and applied a Faraday correction in the RM-range from $-500\,000$ to $+500\,000$\,rad\,m$^{-2}$ with steps of 40\,rad\,m$^{-2}$ using the complex linear polarisation form, $L = (Q + iU) \times e^{-2i\,\textrm{RM}\,\lambda^2}$, where $\lambda$ is the observed wavelength and $Q,U$ are the Stokes $Q$ and $U$ vectors.
Then we performed the frequency summation of the magnitude of $L$ that was written to disk in \textsc{PRESTO} format to be later searched with \textsc{PRESTO}'s \textsc{single\_pulse\_search.py} using boxcar width up to 300 bins and a minimum S/N detection threshold of 7. \citet{lde+21} have argued that searching in linear polarisation with data recorded with a linear feed is still plagued with the red noise fluctuations as in the total intensity search. However Faraday correction with large RM values, i.e. $|\textrm{RM}| \gtrsim 10\,000$\,rad\,m$^{-2}$, depolarises the baseline fluctuations, thus making our search insensitive to these fluctuations.
We therefore applied this minimum RM-threshold when producing the summary plot and restricted our search in linear polarisation to the inner pointing on \sga{} where RM is expected to be the largest.


\subsection{Testing the processing pipeline}
\begin{figure*}
  \begin{center}
    \includegraphics[width=\textwidth]{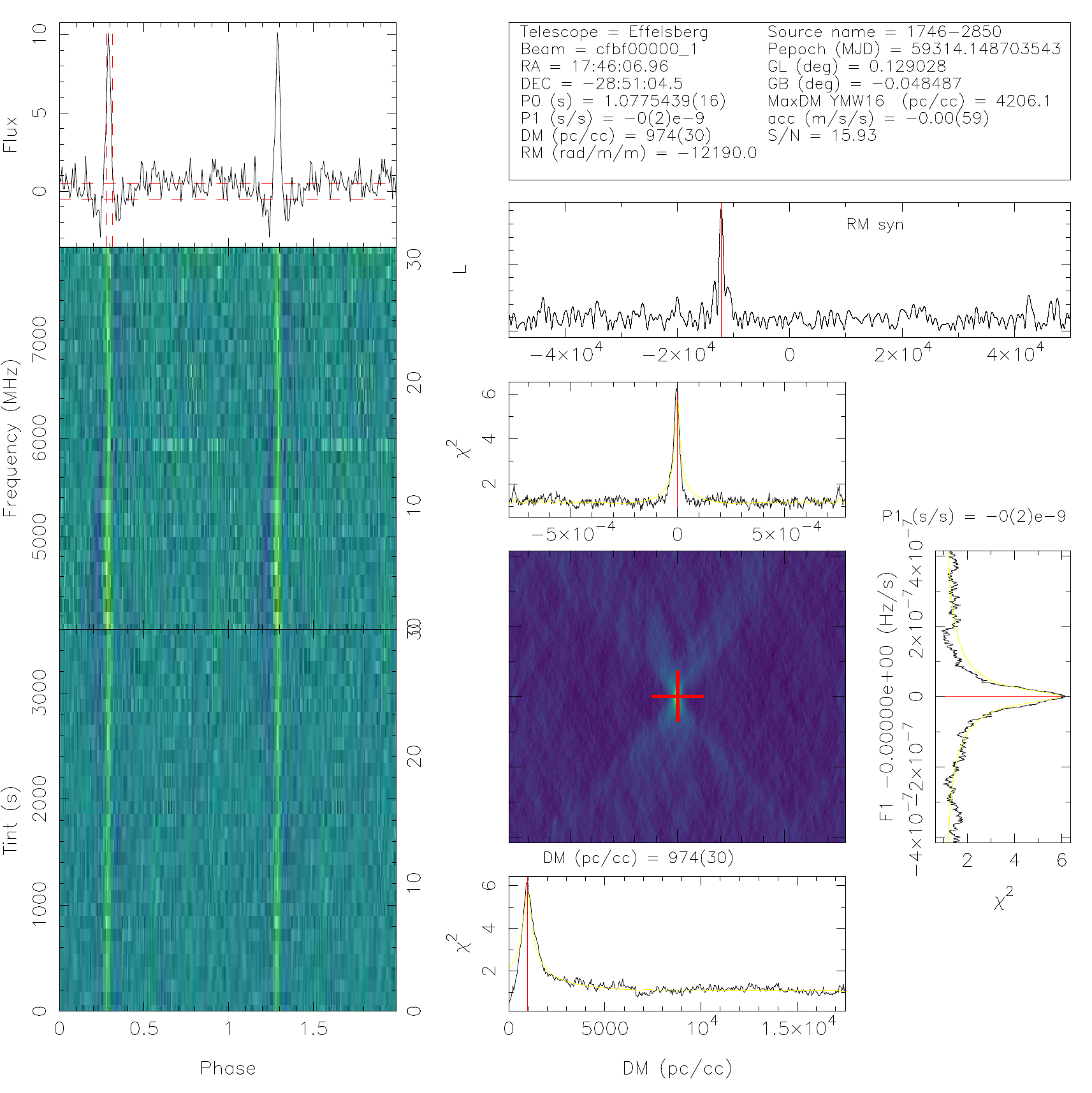}
      \caption{Detection plot of the pulsar J1746-2850 produced by \textsc{PulsarX}. In addition to the panels described in \citet{mbc+23}, a new additional panel (a) shows the results of the RM synthesis analysis of the pulse delimited by the two vertical dashed lines of panel (b) for an evenly spaced grid of RM comprised between -50\,000 rad\,m$^{-2}$ and + 50\,000 rad\,m$^{-2}$ with an RM step of 5\,rad\,m$^{-2}$.}
  \label{fig:1746}
  \end{center}
\end{figure*}

The processing pipeline described above was first tested on an 1-hr long observation of PSR~J1746--2850, recorded as part of the RM monitoring of the GC pulsars with Effelsberg \citep{and+23}, with the same observing setup as the one used in this work. PSR~J1746--2850 \citep{dcl09} is located about 12 arcmin away from \sga{}, close to the Quintuplet cluster, within the Radio Arc Bubble \citep{scc+07}. It has a rotational period of 1.07\,s, a DM of about 960\,pc\,cm$^{-3}$, an RM of $-12\,234\pm181$\,rad\,m$^{-2}$ \citep{and+23} and a pulse profile that is roughly 50\% linearly polarised.
We successfully redetected PSR~J1746$-$2850 in our low acceleration search pipeline with a S/N of 16. The RM estimated from the detected pulse window is $-12\,190$\,rad\,m$^{-2}$ (see Fig.~\ref{fig:1746} for the detection plot) consistent with the measurement reported by \citet{and+23}.

\begin{figure}
  \begin{center}
    \includegraphics[width=\columnwidth]{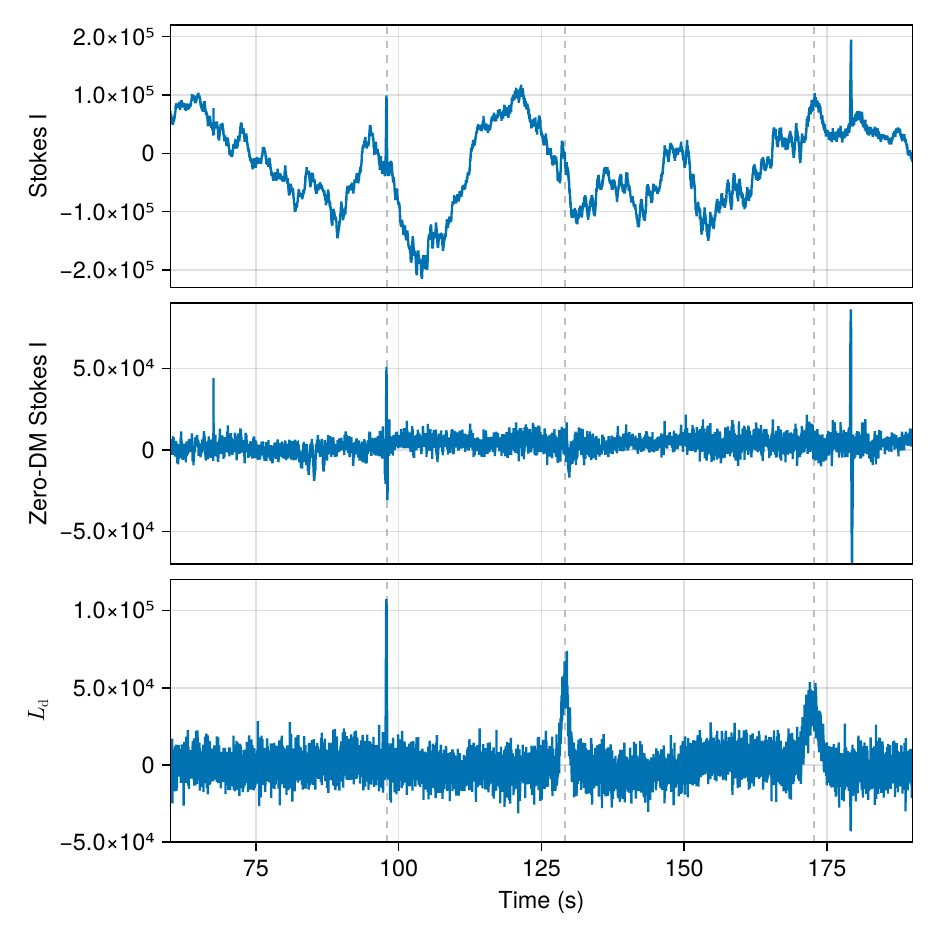}
      \caption{Top panel: an excerpt of the total intensity (Stokes I) time series dedispersed at the DM of the injected pulses and exhibiting large amount of red noise. Middle panel: a view of the Stokes I time series dedispersed at the DM of the injected pulses and after subtraction of the zero DM time series. Bottom panel: time series of the linear polarisation  $L_\textrm{d}$ after dedispersion and the correction of Faraday rotation at $\textrm{RM}=10\,000$\,rad\,m$^{-2}$. For clarity, the average value of $L_\textrm{d}$ has been subtracted from it. In all three panels, the grey dashed lines indicate the location of the three injected pulses. }
  \label{fig:simu}
  \end{center}
\end{figure}

To test the pipeline for slow-transient search in linear polarisation, we injected into one of our calibrated \textsc{PSRFITS} data three simulated full-Stokes pulse profiles created with \textsc{PSRCHIVE} that are fully linearly polarised with an RM of 10\,000\,rad\,m$^{-2}$. These simulated pulse profiles of Gaussian shape share the same DM of 1750\,pc\,cm$^{-3}$ but have widths at a 50\% intensity level of 0.2, 1.5 and 3\,s, respectively. The dedispersed Stokes I time series with and without zero-DM subtraction are shown in Fig.~\ref{fig:simu} along with the 
Faraday rotation corrected
linear  polarisation time series ($L_\textrm{d}$). We ran \textsc{single\_pulse\_search.py} on the three time series and found that only the $L_\textrm{d}$ dataset successfully recovered all three injected pulses without any other spurious candidates with a S/N above 7. This demonstrates that polarised bursts with wide widths can  be efficiently detected in our survey through the search in linear polarisation.

\section{Results and discussion}
\label{sec:res}

\subsection{Survey sensitivity}
We compute the minimum mean flux density $S_\textrm{min}$ of a pulsar with spin period $P_\textrm{s}$ that can be detected by our survey with a significance $\sigma$ using the modified radiometer equation from \citet{cc97},

\begin{equation}
S_\textrm{min} = \frac{\beta \, \sigma\, S_\textrm{sys}} {\sqrt{n_\textrm{p} \, \Delta f \, T_\textrm{int}}} \left( \frac{\sqrt{1-\pi/4} \sqrt{N_\textrm{h}}} {\sum_{h=1}^{N_\textrm{h}}{R(h)}} \right),
\end{equation}
with $S_\textrm{sys} = T_\textrm{sys} / G$ being the System Equivalent Flux Density (SEFD).
Here $\beta \approx 1$ is the degradation factor due to the 8-bit initial digitisation of the data, $G=1.5$\,K\,Jy$^{-1}$ is the average gain of the Effelsberg telescope over our receiver bandwidth and $T_\textrm{sys}$ is the total system noise temperature. The number of summed polarisations $n_\textrm{p}=2$, the chosen number of summed harmonics $N_\textrm{h}$, the effective bandwidth $\Delta f \sim 3400$\,MHz after RFI cleaning and the integration time $ T_\textrm{int}$ are parameters of the survey. The amplitude ratio in the Fourier domain of the h-th harmonic is $R(h)=e^{-(\pi \epsilon h/2 \sqrt{\ln{2}})^2}$, see also e.g. \citet{tle+23} for more details.A sur The duty cycle of a pulsar's averaged pulse, denoted $\epsilon = W_\textrm{eff} / P_\textrm{s}$ with $W_\textrm{eff}$ being the effective pulse width, is affected by the sampling time, the intra-channel dispersion and temporal scattering that broadens the intrinsic pulse width $W_\textrm{i}$ such that
\begin{equation}
  W_\textrm{eff}^2 = W_\textrm{i}^2  +  \delta t^2 + 
  \left( 2\kappa \frac{  \textrm{DM}\, \delta f}{ \nu^3} \right)^2 + \left(\tau_\textrm{1\,GHz} \, \nu^{\alpha} \right)^{2}.
\end{equation}
$\kappa=4.1488 \times 10^3 $\,MHz$^2$\,pc$^{-1}$\,cm$^3$\,s is known as  the dispersion constant, $\delta f \simeq 0.97$\,MHz is the width of a single frequency channel, $\tau_\textrm{1\,GHz}$ is the scattering timescale at a reference frequency $\nu$ of 1\,GHz. For the inner pointing centered on the GC magnetar, we took $\tau_\textrm{1\,GHz}=1.3$\,s \citep{sle+14}. For all other pointings and without prior information, we assumed $\tau_\textrm{1\,GHz}=0.5$\,s. In all cases, the scattering index is $\alpha=-3.8$ \citep{sle+14}.

As reported previously by \citet[e.g.][]{jkl+06,etd+21} thermal and non-thermal emission from the GC dominate the contribution to $T_\textrm{sys}$  at frequencies $ \lesssim 8$\,GHz.  $T_\textrm{sys}$ also includes elevation-dependent contribution from spillover radiation from the ground \citep[see measurements in][]{etd+21}.
Continuum maps of the GC region  \citep[see e.g.][]{srw+89,rfr+90,lyc+08} shows that the continuum emission from the GC at a frequency of 5\,GHz  can reach up to a few hundreds of Jy at the position of \sga and extend several arcmins away from the position of \sga with a contribution of a few tens of Jy.  This contradicts the assumption used in \citet{scc+21} to neglect the GC background contribution in the sensitivity calculations of their outer pointings.

\begin{figure}
  \begin{center}
    \includegraphics[width=\columnwidth]{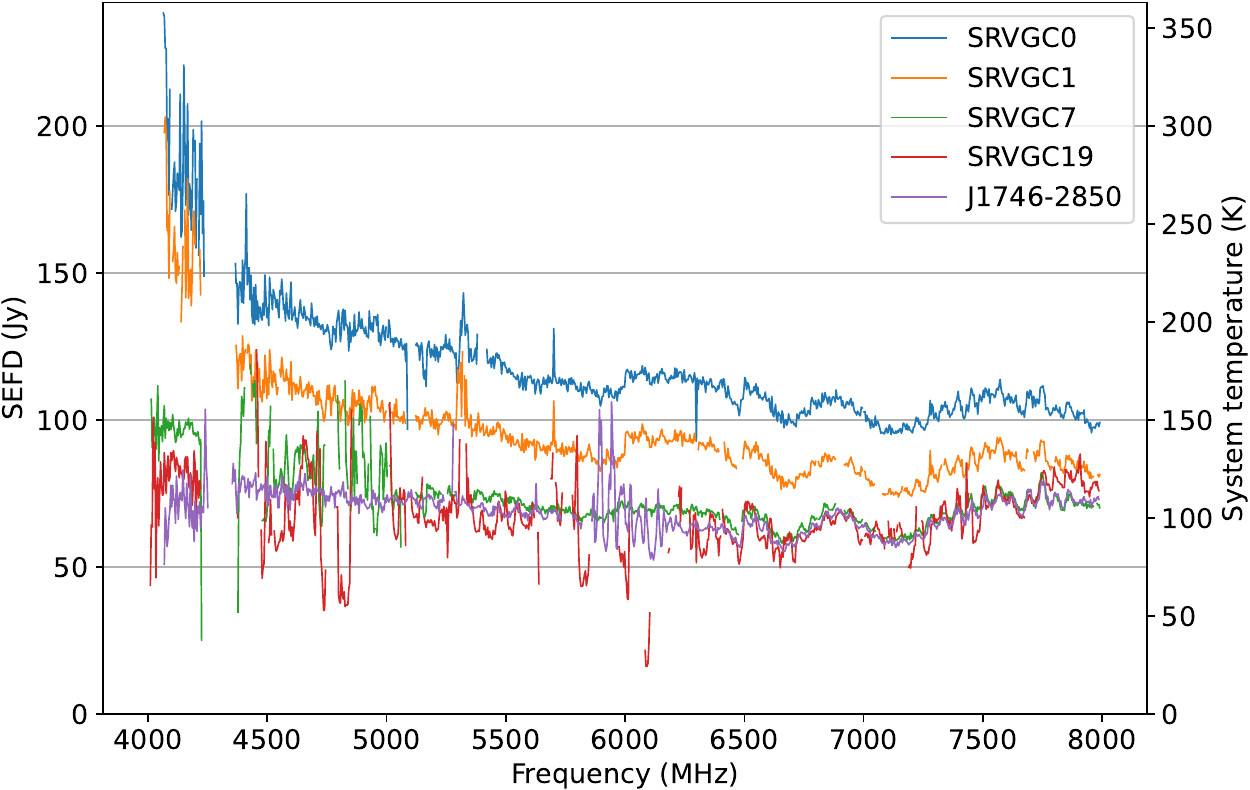}
      \caption{Measured SEFD and $T_\textrm{sys}$ as a function of observing frequency for a set of 4 survey pointings, from inner to outer rings (see Fig.\,\ref{fig:grid}). For comparison, we also added the SEFD measured at the sky location of PSR J1746--2850.  The discontinuities in the SEFD curves are due to zapped frequency channels. }
  \label{fig:Tsys}
  \end{center}
\end{figure}

To properly account for the the GC contribution to our survey sensitivity, we used the individual pulsed noise diode observations fired at the sky position of each pointing and our observations of NGC\,7027 to derive the SEFD for all pointings. SEFD measurements from inner to outer pointings are shown in Fig.~\ref{fig:Tsys} as well as with the SEFD measured at the sky location of PSR~J1746$-$2850.
We found an average SEFD value  of 112\,Jy (equivalent to $T_\textrm{sys}=168$\,K) for the inner pointing that encompasses \sga{} and the GC magnetar J1745$-$2900. This result is consistent with the results from \citet{etd+21} when averaging their measurements at 4.85\,GHz and 8.35\,GHz. For the three outer rings, from inner to outer rings, we measured an SEFD average value of  87, 72, 68\,Jy, respectively. In addition, we measured an average SEFD value towards PSR~J1746$-$2850 similar to the value for 
the outermost
ring. This is again consistent with the continuum map from \citet{lyc+08} as PSR~J1746$-$2850 is located towards the Radio Arc.

\begin{figure}
  \begin{center}
    \includegraphics[width=\columnwidth]{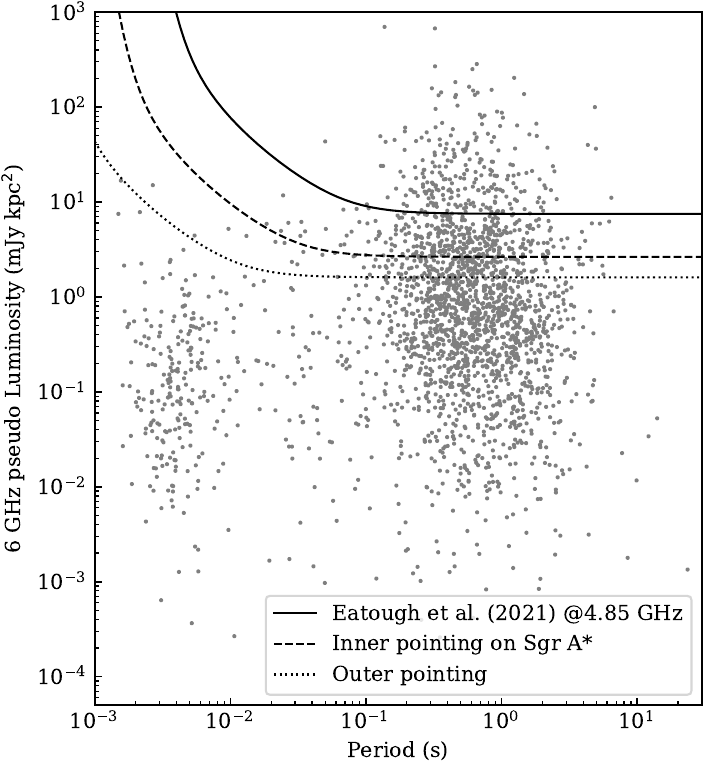}
      \caption{Sensitivity curves of our GC survey as a function of pulsar spin period. The dashed and dotted lines indicate our sensitivity limits for the inner and most outer pointings of our grid with the grey points representing the inferred  pseudo-luminosity of a galactic pulsar population at 6\,GHz. For comparison, we added the well estimated sensitivity limit (the plain line) derived from the \sga{} survey at 4.85\,GHz by \citet{etd+21}.}
  \label{fig:sens}
  \end{center}
\end{figure}

Assuming a $\sigma$ detection threshold of 8 and an intrinsic pulse duty cycle $\epsilon_\textrm{i}=W_\textrm{i}/P_\textrm{s}$ of 5\%, we can derive our sensitivy limits in terms of pulsar luminosity and plot these curves as a function of spin period $P_\textrm{s}$ in Fig.~\ref{fig:sens}. For pulsars with $P_\textrm{s} \gtrsim 50$\,ms, we obtain sensitivity limits between 2.8 and 1.7\,mJy\,kpc$^2$, from the inner pointing on \sga{} towards the outer ring. Below these spin period, scattering effects degrade our survey sensitivity. These limits are significantly higher than the values reported in the GC survey by \citet{scc+21} done at a similar frequency with luminosity of 0.9 and 0.5\,mJy\,kpc$^2$, for the inner pointing on \sga{} and all other pointings, respectively. However we note here that \citet{scc+21} did neglect the GC continuum contribution to all off-\sga{} pointings and used a continuum emission model  \citep{rla17} derived from  maps produced by \citet{lyc+08}

Now, we can estimate the fraction of the known Galactic pulsar population  with a reported flux at 1.4\,GHz (2350 pulsars in the ATNF Pulsar Catalog v\,2.6.2\footnote{https://www.atnf.csiro.au/research/pulsar/psrcat/}) that would be detected by our survey, if placed at the distance of the GC. Following \citet{jvk+18}, we drew 2350 spectral indices from a normal distribution of mean -1.6 with standard deviation of 0.54 to derive the pulsar flux at 6\,GHz. 
We find that not even the currently known most luminous MSPs would be detected if located in the inner pointing on \sga{} while just some of the brightest MSPs could be detected in the outer pointings. Regarding the normal pulsar population (defined as $P_\textrm{s}>30$ ms), up to 30\% of this population could be detected by our survey assuming the best hypothesis (i.e. a pulsar located on-axis of the outer beam). This percentage decreases to 10\% if we assume the sensitivity of a pulsar located on the edge of the inner beam.
It could decrease even further as red noise in the data has been shown to reduce a  survey's sensitivy towards long period pulsars \citep{lbh+15}.
We note here that our observing grid has been tighly packed  assuming the HPBW at our highest observing frequency of 8\,GHz, guaranteeing a more uniform sensitivity compared to past surveys.

\subsection{Sensitivity to orbital motion}
The PRESTO binary pulsar search routine {\sc accelsearch} performs matched filtering to remove Doppler induced frequency drifts in the Fourier detection spectrum. Following \citet{ransom01} and \citet{ar18}, the frequency drift (expressed in Fourier bins) of pulsars under a constant acceleration, and constant jerk, are given by $z=afT_\textrm{int}^2/c$ and $w=jfT_\textrm{int}^3/c$ respectively. Here $f=1/P_\textrm{s}$ is the spin frequency (or chosen harmonic of the spin frequency), $a$ is the line of sight (l.o.s) acceleration, $j$ is the l.o.s jerk and $c$ is the speed of light. \citet{etd+21,lde+21}; and \citet{scc+21} already discussed in detail the impact of orbital motion on the search sensitivity for pulsars orbiting \sga{}. Here our discussion primarily concerns the sensitivity to compact binary systems, although following Section~4.2 of \citet{etd+21} and given our maximum $z$-value of $z_\textrm{max}=200$ bins, we expect to maintain a high degree of sensitivity to spin frequencies up to 1000\,Hz from pulsars with circular orbital periods, $P_{\rm b}\gtrsim\,$120\,d around \sga{} in the central pointing. 

\begin{figure}
  \begin{center}
    \includegraphics[width=\columnwidth]{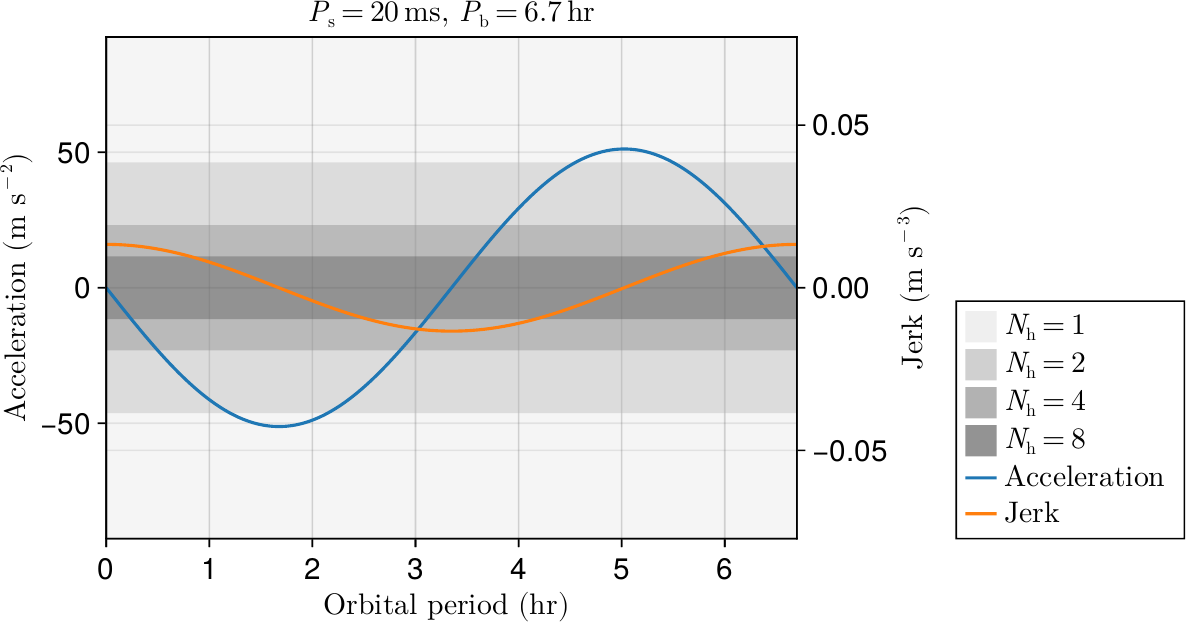}
      \caption{Acceleration (blue line) and jerk (orange line) as a function of time for a binary pulsar of spin period 20\,ms in a 6.7-hr circular orbit around another neutron star of mass 1.35\,M$_\odot$. This simulation assumes an inclination angle of $60$\,degrees. The intersection of the grey bands indicate the sections of the orbit where our \textsc{PRESTO} acceleration and jerk search remains sensitive to the pulsar signal with the specified number of harmonics $N_\textrm{h}$.  }
  \label{fig:acc}
  \end{center}
\end{figure}

Constant acceleration searches are sensitive to binaries where $T_{\rm obs}\lesssim 0.1P_{\rm b}$, and acceleration plus jerk searches where $T_{\rm obs}\sim 0.05P_{\rm b}-0.15P_{\rm b}$ \citep{ransom01,ar18}. 
Therefore, with observations of $T_\textrm{int}=1\,{\rm h}$ it is expected that our survey remains sensitive to minimum orbital periods of  $P_{\rm b}\sim6.7\,{\rm h}$  with our  acceleration plus jerk search. If we consider a pulsar with spin frequency of 50\,Hz in orbit around a second neutron star, we want to establish here if our chosen search range in $z$ and $w$ is sufficient for this detection. 
Fig.\,\ref{fig:acc} shows the fraction of the 6.7\,hr-long orbit where our search can recover the drifted signal of the pulsar up to the desired number of harmonics. We estimate that only 15 and 30\% of the orbit give acceleration low-enough such that the signal can be recovered with up to 8 and 4 harmonics, respectively, potentially severely degrading our sensitivity towards this tight binary. Our $w$-range is enough to cover the full orbit for frequencies up to the 4-th harmonic of the pulsar spin period.

\begin{figure}
  \begin{center}
    \includegraphics[width=\columnwidth]{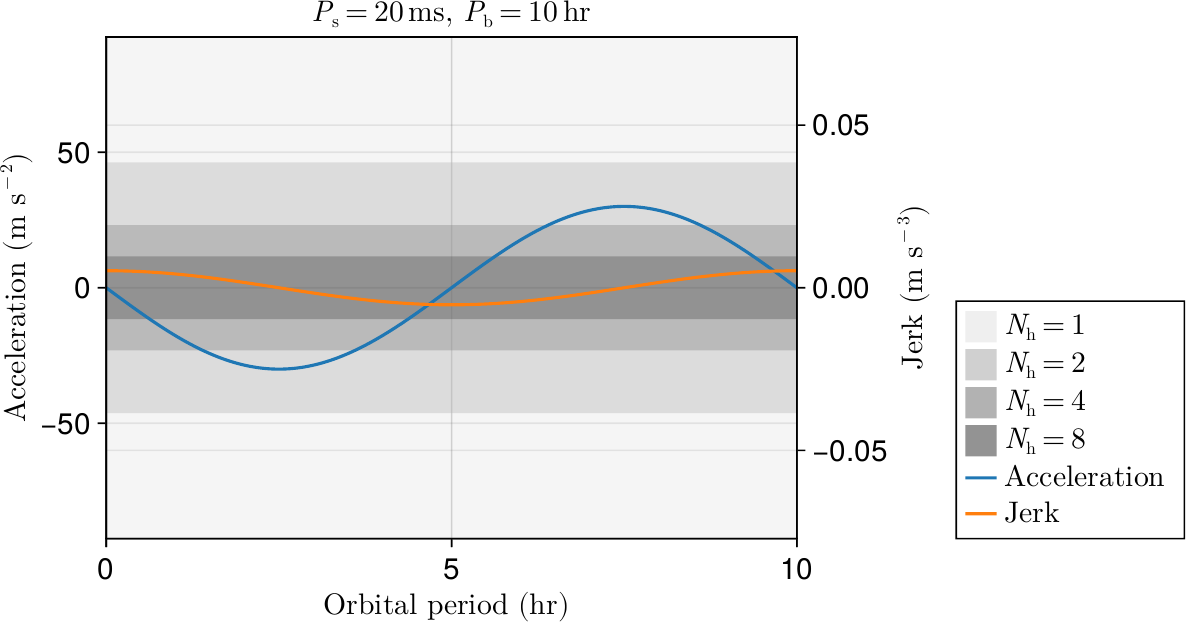}
      \caption{Same as Fig.\ref{fig:acc} but assuming a 10-hr circular orbit.}
  \label{fig:acc2}
  \end{center}
\end{figure}

Considering now a wider $P_{\rm b}\sim10\,{\rm h}$ orbit (see Fig.\,\ref{fig:acc2}), the frequency drift due to acceleration can be recovered in 25\%, 56\%, 100\% of the orbit with up to 8, 4 and 2 harmonics, respectively. In this case, the $w$-range is  sufficient to recover the jerk through the full orbit up to the 8-th harmonic of the pulsar's spin period.

\subsection{Outcome of the survey}

We visually inspected all periodicity candidates produced by \textsc{PulsarX} and found no new pulsar to date. The GC magnetar SGR~J1745$-$2900 was not detected in the periodicity search of the inner pointing with the data recorded on March 2nd, 2020.
The \textsc{TransientX} search for single pulses produced a set of about 2500 candidates and inspection plots from the 37 pointings. Among these candidates, about a tenth were attributed to the GC magnetar SGR~1745$-$2900, with 90\% of them coming from the inner pointing. No promising candidate single pulse has been found with S/N above 8 and we report here on one tentative detection of a narrow pulse at $\textrm{DM}=4550$\,pc\,cm$^{-2}$ with a S/N  above 7 (see Fig.~\ref{fig:SP_cand_SRVGC25}).

\begin{figure*}
  \begin{center}
    \includegraphics[width=\textwidth]{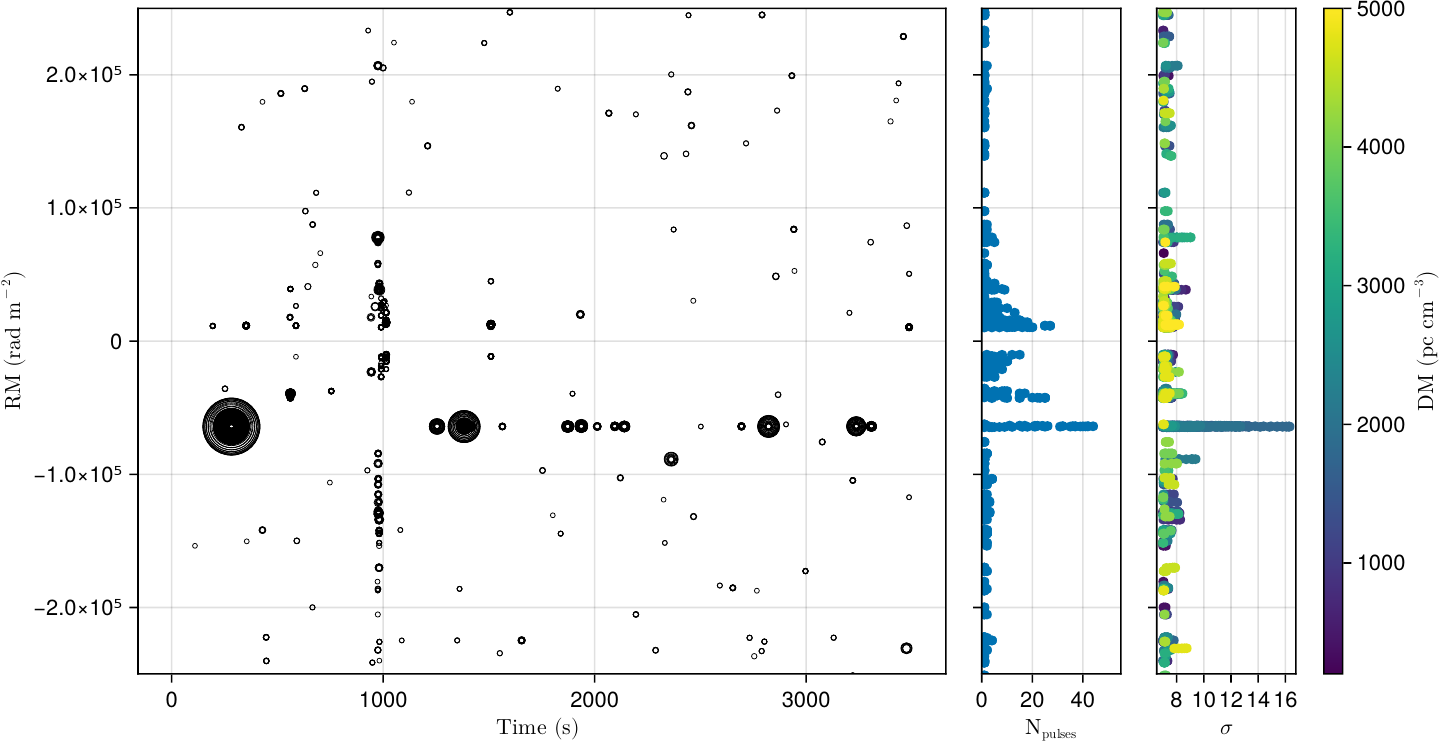}
      \caption{Summary plot of the linear polarisation search. The left panel shows the RM of the single pulse candidates versus time with the size of the open circle marker indicating larger S/N ($\sigma$) with larger marker. The middle panel shows the number of single pulse detection per bin of RM. The panel on the right shows $\sigma $ of the burst as function of its RM with a color scale corresponding to the detection DM. For clarity, the RM-range (y-axis) shown on all panels is restricted to $-250,000;+250,000$\,rad\,m$^{-2}$. As described in Section \ref{sec:ssearch}, candidates with |RM| $<$10,000\,rad\,m$^{-2}$ are excised. The detection of several bright pulses at DM of 1800\,pc\,cm$^{-3}$ and peaking at RM=$-$66,000\,rad\,m$^{-2}$ corrrespond to single pulses emitted by the SGR~J1745$-$2900.}
  \label{fig:SP_L}
  \end{center}
\end{figure*}

After excluding several two-seconds time windows with $\sim 5$ min periodicity that showed a large number of events with wide ranging RMs, the slow transient search in linear polarisation resulted in a set of 1033 pulses with S/N $> 7$ as reported by \textsc{single\_pulse\_search.py} (Fig.~\ref{fig:SP_L}).  The spurious candidates were attributed to a weather radar emitting polarised radio waves around 5.6\,GHz in Germany. All the bright (i.e. with S/N $> 10$) SPs are clustered around $\textrm{DM}=1800$\,pc\,cm$^{-3}$ and RM$=-64000$\,rad\,m$^{-2}$ and most likely originate from the magnetar SGR J1745$-$2900. No other promising candidate was found.

\section{Conclusions}

We conducted between 2019 and 2020 a survey for pulsars and transients within the 10\,pc region centred around \sga{} at a central frequency of 6 GHz with the Effelsberg radio telescope. The survey's sensitivity was maximised by building a compact observing grid of 37 pointings around the sky location of SGR J1745$-$2900.
We analysed the calibrated full-Stokes data with acceleration plus jerk search techniques such that our survey remained sensitive to non-fully recycled MSPs (i.e. with $P_\textrm{s} \sim 20$\,ms) in a putative 10-hr long circular orbit. 
The polarisation information provided another  scoring input in the manually inspected diagnostic plots of pulsar and single pulse candidates produced by \textsc{PulsarX} and \textsc{TransientX}, respectively. No new pulsar or transient pulsed emission have been detected so far.

For the inner pointing encompassing \sga{}, we performed an additional search for slow transient emission in the defaraded linear polarisation. This analysis was justified by the presence of strong red noise in the data inherent to the nature of our single dish  observations of the GC and it is shown to dramatically reduce the number of false alarm events in the single pulse diagnostic plot. However this analysis only redetected the bright single pulse emission of the magnetar SGR~J1745$-$2900.

A careful estimate of our survey sensitivity shows that, although it is a factor of three
more sensitive
than the limit presented by \citet{etd+21}, it is still not sensitive to a population of MSPs in the vicinity of \sga{} and hampered by the strong GC background emission. We also demonstrated that previous sensitivity limits for pulsar searches in the wider GC region were greatly overestimated and that the background noise contribution must be properly assessed as previously reported in \citet{etd+21}. Yet, radio interferometers such as MeerKAT \citep{pbs+23} or the ngVLA \citep{bcc+18} have the ability to resolve the background emission from the \sga{} complex, and to allow for the subtraction of the incoherent beam and thus show exciting prospects for future GC pulsar surveys.

\begin{acknowledgements}
Based on observations with the 100-m telescope of the Max-Planck-Institut für Radioastronomie at Effelsberg. We acknowledge the use of the HPC systems Hercules and Raven at the Max Planck Computing and Data Facility where all computations were performed. The authors gratefully acknowledge support from the  European Research Council (ERC) Synergy Grant “BlackHoleCam” Grant Agreement Number 610058. R.P.E. is supported by the Chinese Academy of Sciences President’s International Fellowship Initiative, grant No. 2021FSM0004.
\end{acknowledgements}

\bibliographystyle{aa}
\bibliography{1810,GC}

\begin{thebibliography}{49}
\expandafter\ifx\csname natexlab\endcsname\relax\def\natexlab#1{#1}\fi

\bibitem[{{Abbate} {et~al.}(2023){Abbate}, {Noutsos}, {Desvignes}, {Wharton},
  {Torne}, {Kramer}, {Eatough}, {Karuppusamy}, {Liu}, {Shao}, \&
  {Wongphechauxsorn}}]{and+23}
{Abbate}, F., {Noutsos}, A., {Desvignes}, G., {et~al.} 2023, \mnras, 524, 2966

\bibitem[{{Andersen} \& {Ransom}(2018)}]{ar18}
{Andersen}, B.~C. \& {Ransom}, S.~M. 2018, \apjl, 863, L13

\bibitem[{{Bower} {et~al.}(2018){Bower}, {Chatterjee}, {Cordes}, {Demorest},
  {Deneva}, {Dexter}, {Kramer}, {Lazio}, {Ransom}, {Shao}, {Wex}, \&
  {Wharton}}]{bcc+18}
{Bower}, G.~C., {Chatterjee}, S., {Cordes}, J., {et~al.} 2018, in Astronomical
  Society of the Pacific Conference Series, Vol. 517, Science with a Next
  Generation Very Large Array, ed. E.~{Murphy}, 793

\bibitem[{{Brentjens} \& {de Bruyn}(2005)}]{bd05}
{Brentjens}, M.~A. \& {de Bruyn}, A.~G. 2005, \aap, 441, 1217

\bibitem[{{Caleb} {et~al.}(2022){Caleb}, {Heywood}, {Rajwade}, {Malenta},
  {Stappers}, {Barr}, {Chen}, {Morello}, {Sanidas}, {van den Eijnden},
  {Kramer}, {Buckley}, {Brink}, {Motta}, {Woudt}, {Weltevrede}, {Jankowski},
  {Surnis}, {Buchner}, {Bezuidenhout}, {Driessen}, \& {Fender}}]{chr+22}
{Caleb}, M., {Heywood}, I., {Rajwade}, K., {et~al.} 2022, Nature Astronomy, 6,
  828

\bibitem[{{Cordes} \& {Chernoff}(1997)}]{cc97}
{Cordes}, J.~M. \& {Chernoff}, D.~F. 1997, \apj, 482, 971

\bibitem[{{Cordes} \& {Lazio}(2002)}]{cl02}
{Cordes}, J.~M. \& {Lazio}, T.~J.~W. 2002, ArXiv Astrophysics e-prints
  [\eprint{arXiv:astro-ph/0207156}]

\bibitem[{Cordes \& {McLaughlin}(2003)}]{cm03}
Cordes, J.~M. \& {McLaughlin}, M.~A. 2003, \apj, 596, 1142

\bibitem[{{Deneva} {et~al.}(2009){Deneva}, {Cordes}, \& {Lazio}}]{dcl09}
{Deneva}, J.~S., {Cordes}, J.~M., \& {Lazio}, T.~J.~W. 2009, \apjl, 702, L177

\bibitem[{{Dexter} \& {O'Leary}(2014)}]{do14}
{Dexter}, J. \& {O'Leary}, R.~M. 2014, \apjl, 783, L7

\bibitem[{{Eatough} {et~al.}(2013){Eatough}, {Falcke}, {Karuppusamy}, {Lee},
  {Champion}, {Keane}, {Desvignes}, {Schnitzeler}, {Spitler}, {Kramer},
  {Klein}, {Bassa}, {Bower}, {Brunthaler}, {Cognard}, {Deller}, {Demorest},
  {Freire}, {Kraus}, {Lyne}, {Noutsos}, {Stappers}, \& {Wex}}]{efk+13}
{Eatough}, R.~P., {Falcke}, H., {Karuppusamy}, R., {et~al.} 2013, \nat, 501,
  391

\bibitem[{{Eatough} {et~al.}(2009){Eatough}, {Keane}, \& {Lyne}}]{ekl09}
{Eatough}, R.~P., {Keane}, E.~F., \& {Lyne}, A.~G. 2009, \mnras, 395, 410

\bibitem[{{Eatough} {et~al.}(2021){Eatough}, {Torne}, {Desvignes}, {Kramer},
  {Karuppusamy}, {Klein}, {Spitler}, {Lee}, {Champion}, {Liu}, {Wharton},
  {Rezzolla}, \& {Falcke}}]{etd+21}
{Eatough}, R.~P., {Torne}, P., {Desvignes}, G., {et~al.} 2021, \mnras, 507,
  5053

\bibitem[{{Faucher-Gigu{\`e}re} \& {Loeb}(2011)}]{fl11}
{Faucher-Gigu{\`e}re}, C.-A. \& {Loeb}, A. 2011, \mnras, 415, 3951

\bibitem[{{GRAVITY Collaboration} {et~al.}(2020){GRAVITY Collaboration},
  {Abuter}, {Amorim}, {Baub{\"o}ck}, {Berger}, {Bonnet}, {Brandner}, {Cardoso},
  {Cl{\'e}net}, {de Zeeuw}, {Dexter}, {Eckart}, {Eisenhauer}, {F{\"o}rster
  Schreiber}, {Garcia}, {Gao}, {Gendron}, {Genzel}, {Gillessen}, {Habibi},
  {Haubois}, {Henning}, {Hippler}, {Horrobin}, {Jim{\'e}nez-Rosales}, {Jochum},
  {Jocou}, {Kaufer}, {Kervella}, {Lacour}, {Lapeyr{\`e}re}, {Le Bouquin},
  {L{\'e}na}, {Nowak}, {Ott}, {Paumard}, {Perraut}, {Perrin}, {Pfuhl},
  {Rodr{\'\i}guez-Coira}, {Shangguan}, {Scheithauer}, {Stadler}, {Straub},
  {Straubmeier}, {Sturm}, {Tacconi}, {Vincent}, {von Fellenberg}, {Waisberg},
  {Widmann}, {Wieprecht}, {Wiezorrek}, {Woillez}, {Yazici}, \&
  {Zins}}]{grav+20}
{GRAVITY Collaboration}, {Abuter}, R., {Amorim}, A., {et~al.} 2020, \aap, 636,
  L5

\bibitem[{{Heywood} {et~al.}(2022){Heywood}, {Rammala}, {Camilo}, {Cotton},
  {Yusef-Zadeh}, {Abbott}, {Adam}, {Adams}, {Aldera}, {Asad}, {Bauermeister},
  {Bennett}, {Bester}, {Bode}, {Botha}, {Botha}, {Brederode}, {Buchner},
  {Burger}, {Cheetham}, {de Villiers}, {Dikgale-Mahlakoana}, {du Toit},
  {Esterhuyse}, {Fanaroff}, {February}, {Fourie}, {Frank}, {Gamatham}, {Geyer},
  {Goedhart}, {Gouws}, {Gumede}, {Hlakola}, {Hokwana}, {Hoosen}, {Horrell},
  {Hugo}, {Isaacson}, {J{\'o}zsa}, {Jonas}, {Joubert}, {Julie}, {Kapp},
  {Kenyon}, {Kotz{\'e}}, {Kriek}, {Kriel}, {Krishnan}, {Lehmensiek},
  {Liebenberg}, {Lord}, {Lunsky}, {Madisa}, {Magnus}, {Mahgoub}, {Makhaba},
  {Makhathini}, {Malan}, {Manley}, {Marais}, {Martens}, {Mauch}, {Merry},
  {Millenaar}, {Mnyandu}, {Mokone}, {Monama}, {Mphego}, {New}, {Ngcebetsha},
  {Ngoasheng}, {Ockards}, {Oozeer}, {Otto}, {Passmoor}, {Patel}, {Peens-Hough},
  {Perkins}, {Ramaila}, {Ramanujam}, {Ramudzuli}, {Ratcliffe}, {Robyntjies},
  {Salie}, {Sambu}, {Schollar}, {Schwardt}, {Schwartz}, {Serylak}, {Siebrits},
  {Sirothia}, {Slabber}, {Smirnov}, {Sofeya}, {Taljaard}, {Tasse}, {Tiplady},
  {Toruvanda}, {Twum}, {van Balla}, {van der Byl}, {van der Merwe}, {Van
  Tonder}, {Van Wyk}, {Venter}, {Venter}, {Wallace}, {Welz}, {Williams}, \&
  {Xaia}}]{hrc+22}
{Heywood}, I., {Rammala}, I., {Camilo}, F., {et~al.} 2022, \apj, 925, 165

\bibitem[{Hu \& Shao(2024)}]{hs24}
Hu, Z. \& Shao, L. 2024, Phys. Rev. Lett., 133, 231402

\bibitem[{{Hurley-Walker} {et~al.}(2022){Hurley-Walker}, {Zhang}, {Bahramian},
  {McSweeney}, {O'Doherty}, {Hancock}, {Morgan}, {Anderson}, {Heald}, \&
  {Galvin}}]{hzb+22}
{Hurley-Walker}, N., {Zhang}, X., {Bahramian}, A., {et~al.} 2022, \nat, 601,
  526

\bibitem[{{Jankowski} {et~al.}(2018){Jankowski}, {van Straten}, {Keane},
  {Bailes}, {Barr}, {Johnston}, \& {Kerr}}]{jvk+18}
{Jankowski}, F., {van Straten}, W., {Keane}, E.~F., {et~al.} 2018, \mnras, 473,
  4436

\bibitem[{{Johnston} {et~al.}(2006){Johnston}, {Kramer}, {Lorimer}, {Lyne},
  {McLaughlin}, {Klein}, \& {Manchester}}]{jkl+06}
{Johnston}, S., {Kramer}, M., {Lorimer}, D.~R., {et~al.} 2006, \mnras, 373, L6

\bibitem[{{Kennea} {et~al.}(2013){Kennea}, {Burrows}, {Kouveliotou}, {Palmer},
  {G{\"o}{\u g}{\"u}{\c s}}, {Kaneko}, {Evans}, {Degenaar}, {Reynolds},
  {Miller}, {Wijnands}, {Mori}, \& {Gehrels}}]{kbk+13}
{Kennea}, J.~A., {Burrows}, D.~N., {Kouveliotou}, C., {et~al.} 2013, \apjl,
  770, L24

\bibitem[{{Klein}(2005)}]{kle05}
{Klein}, B. 2005, PhD thesis, Rheinische Friedrich Wilhelms University of Bonn,
  Germany

\bibitem[{{Kramer} {et~al.}(2000){Kramer}, {Klein}, {Lorimer}, {M{\"u}ller},
  {Jessner}, \& {Wielebinski}}]{kkl+00}
{Kramer}, M., {Klein}, B., {Lorimer}, D., {et~al.} 2000, in Astronomical
  Society of the Pacific Conference Series, Vol. 202, IAU Colloq. 177: Pulsar
  Astronomy - 2000 and Beyond, ed. M.~{Kramer}, N.~{Wex}, \& R.~{Wielebinski},
  37

\bibitem[{Law {et~al.}(2008)Law, Yusef-Zadeh, Cotton, \& Maddalena}]{lyc+08}
Law, C.~J., Yusef-Zadeh, F., Cotton, W.~D., \& Maddalena, R.~J. 2008, The
  Astrophysical Journal Supplement Series, 177, 255

\bibitem[{{Lazarus} {et~al.}(2015){Lazarus}, {Brazier}, {Hessels},
  {Karako-Argaman}, {Kaspi}, {Lynch}, {Madsen}, {Patel}, {Ransom}, {Scholz},
  {Swiggum}, {Zhu}, {Allen}, {Bogdanov}, {Camilo}, {Cardoso}, {Chatterjee},
  {Cordes}, {Crawford}, {Deneva}, {Ferdman}, {Freire}, {Jenet}, {Knispel},
  {Lee}, {van Leeuwen}, {Lorimer}, {Lyne}, {McLaughlin}, {Siemens}, {Spitler},
  {Stairs}, {Stovall}, \& {Venkataraman}}]{lbh+15}
{Lazarus}, P., {Brazier}, A., {Hessels}, J.~W.~T., {et~al.} 2015, \apj, 812, 81

\bibitem[{{Liu} {et~al.}(2021){Liu}, {Desvignes}, {Eatough}, {Karuppusamy},
  {Kramer}, {Torne}, {Wharton}, {Chatterjee}, {Cordes}, {Crew}, {Goddi},
  {Ransom}, {Rottmann}, {Abbate}, {Bower}, {Brinkerink}, {Falcke}, {Noutsos},
  {Hern{\'a}ndez-G{\'o}mez}, {Jiang}, {Johnson}, {Lu}, {Pidopryhora},
  {Rezzolla}, {Shao}, {Shen}, \& {Wex}}]{lde+21}
{Liu}, K., {Desvignes}, G., {Eatough}, R.~P., {et~al.} 2021, \apj, 914, 30

\bibitem[{{Liu} {et~al.}(2012){Liu}, {Wex}, {Kramer}, {Cordes}, \&
  {Lazio}}]{lwk+12}
{Liu}, K., {Wex}, N., {Kramer}, M., {Cordes}, J.~M., \& {Lazio}, T.~J.~W. 2012,
  \apj, 747, 1

\bibitem[{{Lorimer} \& {Kramer}(2004)}]{lk04}
{Lorimer}, D.~R. \& {Kramer}, M. 2004, {Handbook of Pulsar Astronomy, CUP}, ed.
  R.~{Ellis}, J.~{Huchra}, S.~{Kahn}, G.~{Rieke}, \& P.~B. {Stetson}

\bibitem[{{Macquart} {et~al.}(2010){Macquart}, {Kanekar}, {Frail}, \&
  {Ransom}}]{mkf+10}
{Macquart}, J.~P., {Kanekar}, N., {Frail}, D.~A., \& {Ransom}, S.~M. 2010,
  \apj, 715, 939

\bibitem[{{Men} \& {Barr}(2024)}]{mb24}
{Men}, Y. \& {Barr}, E. 2024, arXiv e-prints, arXiv:2401.13834

\bibitem[{{Men} {et~al.}(2023){Men}, {Barr}, {Clark}, {Carli}, \&
  {Desvignes}}]{mbc+23}
{Men}, Y., {Barr}, E., {Clark}, C.~J., {Carli}, E., \& {Desvignes}, G. 2023,
  \aap, 679, A20

\bibitem[{{Padmanabh} {et~al.}(2023){Padmanabh}, {Barr}, {Sridhar}, {Rugel},
  {Damas-Segovia}, {Jacob}, {Balakrishnan}, {Berezina}, {Bernadich},
  {Brunthaler}, {Champion}, {Freire}, {Khan}, {Kl{\"o}ckner}, {Kramer}, {Ma},
  {Mao}, {Men}, {Menten}, {Sengupta}, {Venkatraman Krishnan}, {Wucknitz},
  {Wyrowski}, {Bezuidenhout}, {Buchner}, {Burgay}, {Chen}, {Clark},
  {K{\"u}nkel}, {Nieder}, {Stappers}, {Legodi}, \& {Nyamai}}]{pbs+23}
{Padmanabh}, P.~V., {Barr}, E.~D., {Sridhar}, S.~S., {et~al.} 2023, \mnras,
  524, 1291

\bibitem[{{Psaltis} {et~al.}(2016){Psaltis}, {Wex}, \& {Kramer}}]{pwk16}
{Psaltis}, D., {Wex}, N., \& {Kramer}, M. 2016, \apj, 818, 121

\bibitem[{{Rajwade} {et~al.}(2017){Rajwade}, {Lorimer}, \& {Anderson}}]{rla17}
{Rajwade}, K.~M., {Lorimer}, D.~R., \& {Anderson}, L.~D. 2017, \mnras, 471, 730

\bibitem[{{Ransom}(2001)}]{ransom01}
{Ransom}, S.~M. 2001, PhD thesis, Harvard University, Massachusetts

\bibitem[{{Ransom} {et~al.}(2002){Ransom}, {Eikenberry}, \&
  {Middleditch}}]{rem02}
{Ransom}, S.~M., {Eikenberry}, S.~S., \& {Middleditch}, J. 2002, \aj, 124, 1788

\bibitem[{{Rea} {et~al.}(2013){Rea}, {Esposito}, {Pons}, {Turolla}, {Torres},
  {Israel}, {Possenti}, {Burgay}, {Vigan{\`o}}, {Papitto}, {Perna}, {Stella},
  {Ponti}, {Baganoff}, {Haggard}, {Camero-Arranz}, {Zane}, {Minter},
  {Mereghetti}, {Tiengo}, {Sch{\"o}del}, {Feroci}, {Mignani}, \&
  {G{\"o}tz}}]{rep+13}
{Rea}, N., {Esposito}, P., {Pons}, J.~A., {et~al.} 2013, \apjl, 775, L34

\bibitem[{{Reich} {et~al.}(1990){Reich}, {F{\"u}rst}, {Reich}, \&
  {Reif}}]{rfr+90}
{Reich}, W., {F{\"u}rst}, E., {Reich}, P., \& {Reif}, K. 1990, \aaps, 85, 633

\bibitem[{{Schnitzeler} \& {Lee}(2015)}]{sl15}
{Schnitzeler}, D.~H.~F.~M. \& {Lee}, K.~J. 2015, \mnras, 447, L26

\bibitem[{{Seiradakis} {et~al.}(1989){Seiradakis}, {Reich}, {Wielebinski},
  {Lasenby}, \& {Yusef-Zadeh}}]{srw+89}
{Seiradakis}, J.~H., {Reich}, W., {Wielebinski}, R., {Lasenby}, A.~N., \&
  {Yusef-Zadeh}, F. 1989, \aaps, 81, 291

\bibitem[{{Simpson} {et~al.}(2007){Simpson}, {Colgan}, {Cotera}, {Erickson},
  {Hollenbach}, {Kaufman}, \& {Rubin}}]{scc+07}
{Simpson}, J.~P., {Colgan}, S. W.~J., {Cotera}, A.~S., {et~al.} 2007, \apj,
  670, 1115

\bibitem[{{Spitler} {et~al.}(2014){Spitler}, {Lee}, {Eatough}, {Kramer},
  {Karuppusamy}, {Bassa}, {Cognard}, {Desvignes}, {Lyne}, {Stappers}, {Bower},
  {Cordes}, {Champion}, \& {Falcke}}]{sle+14}
{Spitler}, L.~G., {Lee}, K.~J., {Eatough}, R.~P., {et~al.} 2014, \apjl, 780, L3

\bibitem[{{Suresh} {et~al.}(2022){Suresh}, {Cordes}, {Chatterjee}, {Gajjar},
  {Perez}, {Siemion}, {Lebofsky}, {MacMahon}, \& {Ng}}]{scc+21}
{Suresh}, A., {Cordes}, J.~M., {Chatterjee}, S., {et~al.} 2022, \apj, 933, 121

\bibitem[{{Torne} {et~al.}(2021){Torne}, {Desvignes}, {Eatough}, {Kramer},
  {Karuppusamy}, {Liu}, {Noutsos}, {Wharton}, {Kramer}, {Navarro}, {Paubert},
  {Sanchez}, {Sanchez-Portal}, {Schuster}, {Falcke}, \& {Rezzolla}}]{tde+21}
{Torne}, P., {Desvignes}, G., {Eatough}, R.~P., {et~al.} 2021, \aap, 650, A95

\bibitem[{{Torne} {et~al.}(2023){Torne}, {Liu}, {Eatough}, {Wongphechauxsorn},
  {Cordes}, {Desvignes}, {De Laurentis}, {Kramer}, {Ransom}, {Chatterjee},
  {Wharton}, {Karuppusamy}, {Blackburn}, {Janssen}, {Chan}, {Crew}, {Matthews},
  {Goddi}, {Rottmann}, {Wagner}, {S{\'a}nchez}, {Ruiz}, {Abbate}, {Bower},
  {Salamanca}, {G{\'o}mez-Ruiz}, {Herrera-Aguilar}, {Jiang}, {Lu}, {Pen},
  {Raymond}, {Shao}, {Shen}, {Paubert}, {Sanchez-Portal}, {Kramer}, {Castillo},
  {Navarro}, {John}, {Schuster}, {Johnson}, {Rygl}, {Akiyama}, {Alberdi},
  {Alef}, {Algaba}, {Anantua}, {Asada}, {Azulay}, {Bach}, {Baczko}, {Ball},
  {Balokovi{\'c}}, {Barrett}, {Baub{\"o}ck}, {Benson}, {Bintley}, {Blundell},
  {Bouman}, {Boyce}, {Bremer}, {Brinkerink}, {Brissenden}, {Britzen},
  {Broderick}, {Broguiere}, {Bronzwaer}, {Bustamante}, {Byun}, {Carlstrom},
  {Ceccobello}, {Chael}, {Chang}, {Chatterjee}, {Chen}, {Chen}, {Cheng}, {Cho},
  {Christian}, {Conroy}, {Conway}, {Crawford}, {Cruz-Osorio}, {Cui}, {Dahale},
  {Davelaar}, {Deane}, {Dempsey}, {Dexter}, {Dhruv}, {Doeleman}, {Dougal},
  {Dzib}, {Emami}, {Falcke}, {Farah}, {Fish}, {Fomalont}, {Ford}, {Foschi},
  {Fraga-Encinas}, {Freeman}, {Friberg}, {Fromm}, {Fuentes}, {Galison},
  {Gammie}, {Garc{\'\i}a}, {Gentaz}, {Georgiev}, {Gold}, {G{\'o}mez}, {Gu},
  {Gurwell}, {Hada}, {Haggard}, {Haworth}, {Hecht}, {Hesper}, {Heumann}, {Ho},
  {Ho}, {Honma}, {Huang}, {Huang}, {Hughes}, {Ikeda}, {Impellizzeri}, {Inoue},
  {Issaoun}, {James}, {Jannuzi}, {Jeter}, {Jim{\'e}nez-Rosales}, {Jorstad},
  {Joshi}, {Jung}, {Karami}, {Kawashima}, {Keating}, {Kettenis}, {Kim}, {Kim},
  {Kim}, {Kim}, {Kino}, {Koay}, {Kocherlakota}, {Kofuji}, {Koyama},
  {Krichbaum}, {Kuo}, {La Bella}, {Lauer}, {Lee}, {Lee}, {Leung}, {Levis},
  {Li}, {Lico}, {Lindahl}, {Lindqvist}, {Lisakov}, {Liu}, {Liuzzo}, {Lo},
  {Lobanov}, {Loinard}, {Lonsdale}, {MacDonald}, {Mao}, {Marchili}, {Markoff},
  {Marrone}, {Marscher}, {Mart{\'\i}-Vidal}, {Matsushita}, {Medeiros},
  {Menten}, {Michalik}, {Mizuno}, {Mizuno}, {Moran}, {Moriyama},
  {Moscibrodzka}, {M{\"u}ller}, {M{\"u}ller}, {Mus}, {Musoke}, {Myserlis},
  {Nadolski}, {Nagai}, {Nagar}, {Nakamura}, {Narayan}, {Narayanan},
  {Natarajan}, {Nathanail}, {Neilsen}, {Neri}, {Ni}, {Noutsos}, {Nowak}, {Oh},
  {Okino}, {Olivares}, {Ortiz-Le{\'o}n}, {Oyama}, {{\"O}zel}, {Palumbo},
  {Paraschos}, {Park}, {Parsons}, {Patel}, {Pesce}, {Pi{\'e}tu}, {Plambeck},
  {PopStefanija}, {Porth}, {P{\"o}tzl}, {Prather}, {Preciado-L{\'o}pez},
  {Psaltis}, {Pu}, {Ramakrishnan}, {Rao}, {Rawlings}, {Rezzolla}, {Ricarte},
  {Ripperda}, {Roelofs}, {Rogers}, {Ros}, {Romero-Ca{\~n}izales},
  {Roshanineshat}, {Roy}, {Ruszczyk}, {S{\'a}nchez-Arg{\"u}elles}, {Sasada},
  {Satapathy}, {Savolainen}, {Schloerb}, {Schonfeld}, {Small}, {Sohn},
  {SooHoo}, {Souccar}, {Sun}, {Tetarenko}, {Tiede}, {Tilanus}, {Titus},
  {Toscano}, {Traianou}, {Trent}, {Trippe}, {Turk}, {van Bemmel}, {van
  Langevelde}, {van Rossum}, {Vos}, {Ward-Thompson}, {Wardle}, {Weintroub},
  {Wex}, {Wielgus}, {Wiik}, {Witzel}, {Wondrak}, {Wong}, {Wu}, {Yadlapalli},
  {Yamaguchi}, {Yfantis}, {Yoon}, {Young}, {Young}, {Younsi}, {Yu}, {Yuan},
  {Yuan}, {Zensus}, {Zhang}, {Zhao}, \& {Zhao}}]{tle+23}
{Torne}, P., {Liu}, K., {Eatough}, R.~P., {et~al.} 2023, \apj, 959, 14

\bibitem[{{van Straten} {et~al.}(2012){van Straten}, {Demorest}, \&
  {Oslowski}}]{sdo12}
{van Straten}, W., {Demorest}, P., \& {Oslowski}, S. 2012, Astronomical
  Research and Technology, 9, 237

\bibitem[{{Wharton}(2017)}]{wha17}
{Wharton}, R.~S. 2017, PhD thesis, Cornell University, New York

\bibitem[{{Wharton} {et~al.}(2012){Wharton}, {Chatterjee}, {Cordes}, {Deneva},
  \& {Lazio}}]{wcc+12}
{Wharton}, R.~S., {Chatterjee}, S., {Cordes}, J.~M., {Deneva}, J.~S., \&
  {Lazio}, T.~J.~W. 2012, \apj, 753, 108

\bibitem[{{Zhao} {et~al.}(2016){Zhao}, {Morris}, \& {Goss}}]{zmg16}
{Zhao}, J.-H., {Morris}, M.~R., \& {Goss}, W.~M. 2016, \apj, 817, 171

\end{thebibliography}

\appendix
\section{Inspection plot}
\begin{figure*}
  \begin{center}
    \includegraphics[height=0.9\textwidth,angle=-90]{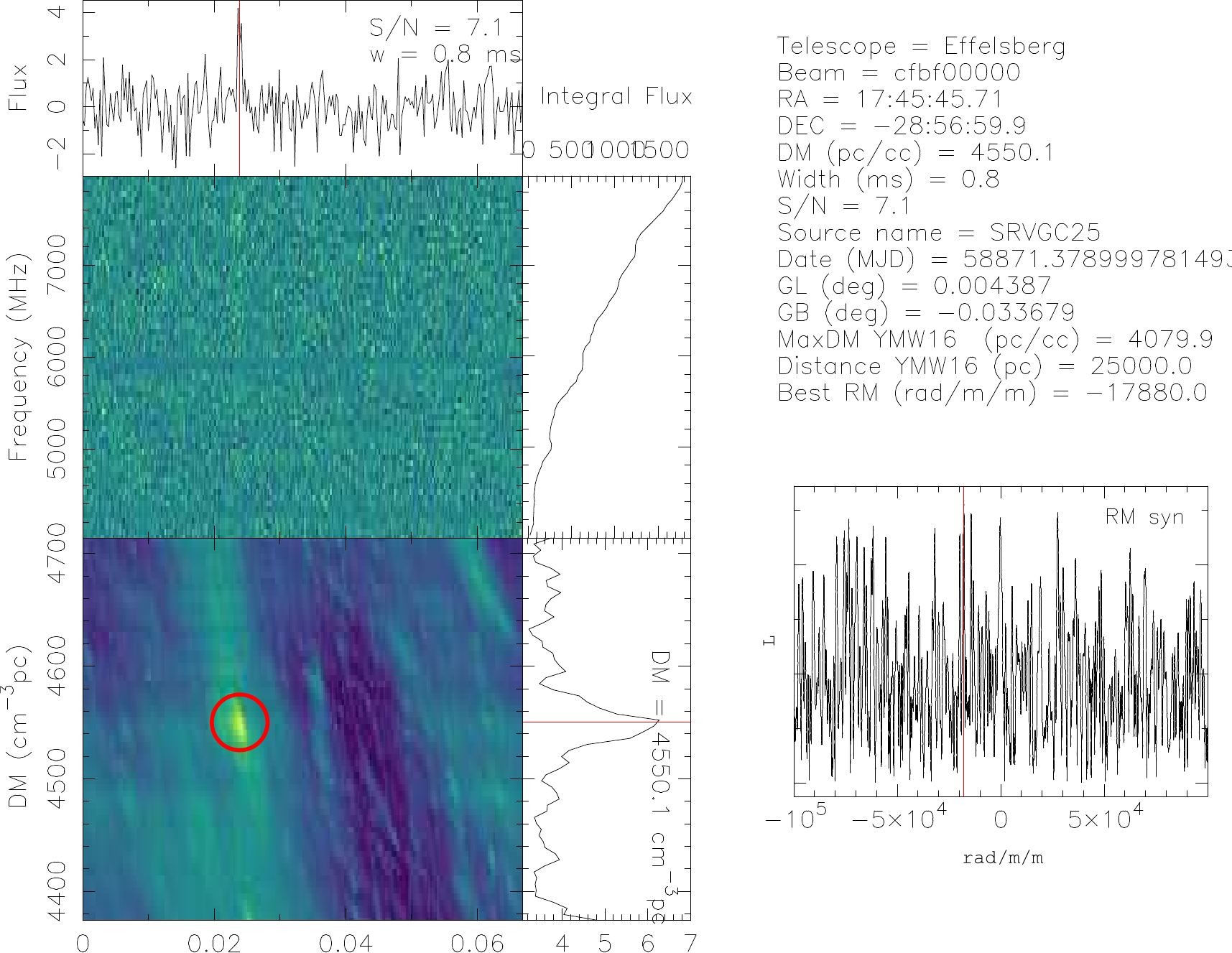}
      \caption{Single pulse candidate produced by \textsc{TransientX}, see \cite{mb24} for a description of the panels. Faraday synthesis is applied to the detected pulse window with the result shown in the bottom-right panel. }
  \label{fig:SP_cand_SRVGC25}
  \end{center}
\end{figure*}

\end{document}